# THE NEED FOR A PARADIGM SHIFT IN OPTICAL ASTRONOMY: A SOLUTION GIVEN BY LIQUID MIRRORS AND EXAMPLES OF THEIR APPLICATIONS TO COSMOLOGY

Ermanno F. Borra, Centre d'Optique, Photonique et Laser,
Observatoire du Mont Mégantic,
Département de Physique, Université Laval, Canada G1K 7P4
Fax: 418-656-2040
Email: borra@phy.ulaval.ca

RECEIVED______________________________________

# ABSTRACT

I argue that there is a crisis in optical Astronomy due to a paucity of telescopes and thus the need for a paradigm shift in telescope technology. Large increases in collecting areas and observing time/astronomer are only possible if we forego the fully steerable multipurpose telescope with a glass primary mirror that has dominated astronomical research. Only by adopting entirely novel technologies that allow one to build large *and inexpensive telescopes* can we achieve truly large improvements. This may come at the expense of versatility and may entail changes in the observing strategies astronomers are now accustomed to. I build my case around a new technology, liquid mirrors, that although in its infancy has achieved credibility. I argue that forthcoming technological improvements will make Liquid Mirror Telescopes (LMTs) nearly as versatile as conventional telescopes. I address the issue of the fields accessible to LMTs equipped with novel optical correctors. Optical design work and exploratory laboratory work indicate that a single LMT should be able to access, with excellent images, subregions anywhere inside fields as large as 45 degrees.

As a practical example of what an LMT can do with the present technology, I examine the expected performance of a 5-m liquid mirror telescope, presently under construction, dedicated to a cosmological survey. It is rather impressive, due to the fact that the instrument works full-time on a four-year survey: Spectrophotometry reaches B=24 with a signal to noise ratio of 10 within a 200Å bandpass for all objects within 300 square degrees and wide-band photometry reaches about B=27. I give three examples of cosmological projects that can be done with the data.

## 1. INTRODUCTION

Astronomical telescopes collect photons, so that their ultimate limitation is set by photon statistics and the standard error $\sigma$ of a measurement is given by $\sqrt{N}$, where $N$ is the total number of photons counted. If we take $S$ to be the photon counts arriving at the earth from an object per unit surface and unit time, $A$ the collecting area of the telescope and $t$ the total integration time, the signal to noise ratio associated with an astronomical observation is given by

$$N/\sigma = 1/\sqrt{ESAt}, \qquad (1)$$

where $E$ is an efficiency factor. If $N$ is the sum of all photons counted over all observations taken by a telescope over a time $t$, Eq. 1 gives the signal to noise ratio of all of the observations taken by that telescope and therefore gives a measure of the information collected by that telescope. We should keep this equation in mind since the theme of this article is that what fundamentally matters is to maximize $At$ by whatever means we can and then try and obtain reasonable values of $E$ by making clever use of the type of telescope and by devising innovative efficient instrumentation. Of course, the actual situation is more complex as one really should take into account the sky background, seeing, detector noise, and other details of the instrumentation. However Eq. 1 is useful since it quantifies a fundamental limit: one can do worse, but not better, than the limitation imposed by photon statistics. One can develop powerful and efficient instruments, devise clever observing strategies, find outstanding sites, so that $E$ nears unity; but one will never do better than predicted by Eq. 1 with $E$=1. Its straightforward prediction being that one should maximize $At$, the obvious conclusion is that to increase our knowledge of the cosmos, one should count more photons and thus increase the total area of the telescopes of the world and the time they spend observing. The type of telescope is somewhat irrelevant. This principle must of course, be sensibly applied: for example building a very large number of very small telescopes is not a sensible approach.

Equation 1 has always guided instrumental developments in Astronomy. In the first half of this century, low quantum efficiency photographic emulsions were the detectors of choice and $E$ was small so that efforts were made to maximize $A$ by building increasingly larger telescopes, culminating with the Hale 5-m telescope which was thought to have reached the limit attainable with a glass mirror. Subsequently, the effort went to increase the quantum efficiencies of detectors, culminating with the development of CCD detectors that reach 80 % quantum efficiencies. Now that detectors approach 100 % efficiencies, the effort is directed again at increasing $A$ by building large telescopes that push the glass mirror technology to its limits. However, this new generation of telescopes increases $A$ by only a factor of about four and are hugely expensive so that few will be built and the World increase of $At$ will not be dramatic. Low cost is an essential ingredient of future mirror technologies, arguably more than large size, since, more telescopes being built, the total area of all of the telescopes of the world will increase with decreasing cost. This sketchy historical review obviously glosses over many instrumental developments such as the advent of fiber spectrographs that allow one to carry out multi object spectroscopy, the advent of Space Astronomy and the opening of new spectral windows.

I argue in this article that there is a need for a paradigm shift in telescope technology and use, and that large increases in $At$ are only possible if we forego the fully steerable multipurpose telescope with a glass primary mirror that has dominated astronomical research. Only by adopting entirely novel mirror technologies that allow one to build large *and inexpensive telescopes* can we achieve truly large increases in $At$. This may come at the expense of versatility and may entail a change in the observing strategies astronomers are now accustomed to. I build my case around a new technology, liquid mirrors, that although in its infancy has achieved credibility. I argue that forthcoming technological improvements will make Liquid Mirror Telescopes (LMTs) nearly as versatile as conventional telescopes. As a practical example, I show how LMTs can, with the present technology, make significant contributions to cosmological research.

## 2. A CRISIS IN OPTICAL ASTRONOMY

In this article I use the term optical Astronomy in a broad sense, spanning the infrared as well as the ultraviolet parts of the spectrum only reachable from space.

Optical Astronomy suffers from a paucity of telescopes. Conventional telescopes are very expensive and can only be justified by sharing them among many investigators. As a consequence, obtaining telescope time is a very competitive process so that only a few privileged astronomers manage to get of the order of 3 nights/year on 4-m class telescopes. The next generation of large telescopes will not noticeably change the situation since they will be few in numbers. The crisis may not be as acutely felt within the handful of institutions that own large telescopes (e.g. Caltech) but it certainly is felt among astronomers in industrialized countries (e.g. Europe or Canada) that have a share in a large telescope. As for the majority of the world, it is essentially left out of competitive astronomical research. Further consider that the average time actually spent observing is of the order of 3.5 hours per night (Benn , & Martin, 1987), the remainder being lost to weather, technical problems and overhead (slewing the telescope, acquiring and identifying the field, reading the data, etc..). The unfortunate consequence is that it typically takes a decade to gather enough data for a substantial observing program, at times rendering its original purpose outdated. Furthermore, successful applicants feel considerable pressure to publish since publication of results obtained from previous observing runs is obviously deemed helpful, if not necessary, to obtain more telescope time. As a consequence all possible information is squeezed out of data, and marginal data (sometime from a back-up program) undergo endless processing to correct for instrumental deficiencies, or low signal. One may argue that this makes an optimal use of expensive telescope time; I however argue that this squanders the most valuable of resources: human time.

Another significant, but hard to quantify, penalty is paid in terms of conservatism in the selection of successful observing programs. Time allocation committees are notoriously conservative and tend to grant observing time to programs most likely to lead to successful publication. This practical approach is imposed by the cost of an observing night; however, by favoring normal research and excluding speculative research, it tends to exclude revolutionary discoveries. Faced with this situation, astronomers naturally tend to be conservative and consider normal research rather than venture into high-risk research. Also, observing programs requiring more than a week observing time per year on large telescopes do not fare well with observing committees and are simply not envisioned by most astronomers, who tend to favor programs that need a small number of nights.

Given these premises, is not surprising that a large fraction of the significant discoveries in optical astronomy in this century was done by Carnegie and Caltech astronomers who had access to a large value of $At$ because they owned the largest telescopes in the world and could obtain several tens of nights a year each. Furthermore, observing committees notwithstanding, they had a large latitude in what they could observe, whether in the original proposal or not, and certainly did not bother with marginal data. My personal value of $At$ was at its highest during my 2-year tenure of a Carnegie Fellowship at the then Hale Observatories(Mt Wilson, Palomar and Las Campanas observatories). It is also not surprising that some of the most exciting discoveries in the second half of the century have come from small telescopes that could be dedicated to long term projects (e.g. the structure seen in the CfA survey; De Lapparent, Geller, & Huchra 1986) or could be used for unusual ideas (e.g. the discovery of very large magnetic fields in white dwarfs; Kemp et al. 1970).

This is a crisis that can be solved by reinventing the telescope and developing technologies that make large *inexpensive* optics. The key word in the previous sentence is: *inexpensive*. Telescopes should be like powerful microscopes: plentiful, relatively inexpensive and easily accessible. Equation 1 shows that the area *A* and the integration time *t* play an equivalent role. The area *A* of a telescope can be increased through some new technology while the time *t* available to an individual observer can be increased by building, for the cost of a classical telescope, several inexpensive innovative telescopes. Low cost is just as important as large size since small institutions can then afford to build their own large instrument that will be shared among few researchers who, individually, have thus a large value of *At*. How large should *A* be? As large as possible, of course; although issues like detector matching and corrector design must be taken into account; but in the 1990s, a 4-m diameter is quite competitive. Presently, only a few wealthy countries can afford such instruments that are few in numbers. The vast majority of countries and institutions cannot afford a share of a 4-m telescope. Into the next century significantly larger diameters should be envisioned, hopefully larger then 30-m so that new technologies should aim for these diameters. Of course such instruments will be expensive, but it is obviously better to spend 100 Megadollars for an array of 30-meters than for an array of 8-meters.

In the next section I argue that considerable savings can be achieved by using fixed telescopes. In the section after, I show that another considerable saving can be achieved by using liquid mirrors.

## 3. FIXED TELESCOPES

considerable savings can be achieved by foregoing steerability since a substantial fraction of the cost of an observatory resides in the telescope mount and the rotating dome. As a matter of fact, the new generation of observatories achieve much of their lower costs by using compact altazimuth mounts and compact "domes". With a fixed telescope, the "dome" is an inexpensive structure. For example the observatory for the UBC-Laval 5-meter telescope presently uder construction has cost $US 60,000, and that comprises dynamiting and extensive foundation works on a mountainous site. The frame of the telescope is also quite inexpensive since it simply is a tower ; unfortunately reduced sky coverage is the price to pay for foregoing steerability. The field corrector then becomes extremely important since the usefulness of a fixed telescope obviously increases with the area of sky that it can access.

The field accessible with corrector designs presently used in astronomical telescopes is of the order of a couple of degrees, allowing a fixed telescope to observe a strip of sky having an area of a few hundred square degrees, thanks to the rotation of the earth around its axis and around the sun. This is actually a large region of sky that can be used for a variety of astronomical studies as shown in section 5, but one can do better. The situation can be considerably improved if we consider novel corrector designs optimized for fixed telescopes. This area of optical design has been neglected since there was little need for it in the era of steerable telescopes, but it has recently received attention, starting with a seminal paper by Richardson and Morbey (1987).

Borra (1993) has recently explored the theoretical limits of correctors for fixed telescopes with a simple argument: Let us consider a zenith telescope observing at $\theta$ degrees from the zenith, and let us assume that a perfect correction is applied onto an image of the pupil of the primary mirror, as is done with adaptive optics. Having perfect correction at the center of the field, the wavefront aberration $\Delta\theta$ arcseconds away is given, to an excellent approximation for $\Delta\theta$ of the order of a few arcseconds, by

$$\Omega(r, \alpha, \theta + \Delta\theta) = \partial W(r, \alpha, \theta)/\partial\theta \; \Delta\theta \; . \qquad (2)$$

where $W(r, \alpha, \theta)$ represents the wavefront, and r and $\alpha$ are the polar coordinates on the mirror. An elaboration of this simple analysis gave the surprising result that the aberrations of a parabola used off-axis can, in principle, be corrected in small patches to zenith distances as high as 45 degrees ($\pm$ 22.5 from zenith). Correcting the pupil is not the best strategy to obtain the widest field of view: This assumption was made simply because an analytical solution could be found and the results of Borra (1993) should only be used as guidelines with the understanding that better corrector designs should be feasible. Indeed, subsequent work has identified two promising designs. First, we have studied a simple system consisting of a single active spherical secondary to which were added aberrations up to 7th order. It gave subarcsecond images in 10X10 arcseconds subregions anywhere within a 20 degree field (Wang, Moretto, Borra, & Lemaître 1994); a performance insufficient for imagery but useful for high-resolution spectroscopy of compact objects. Tracking would be done mechanically, and the corrector warped in real time. Light would be piped to a fixed spectrograph with optical fibers. Adding adaptive optics would render it useful for high-resolution imagery since the field of view would then be limited by the small size of the isoplanatic patch. Experimental work (Moretto et al. 1995) has shown that it is possible to mechanically bend a metallic mirror and add several aberrations to its surface as required: third and fifth - order aberrations were successfully added to a 20-cm diameter mirror.

A second design (Borra, Moretto , & Wang 1995) uses a 2-mirror corrector (dubbed BMW) that gives excellent performance for imagery, yielding encircled energy diameters <0.5 arcseconds in 15 arcminute patches anywhere within a field greater than 45 degrees ($\pm$22.5 degrees from zenith) that makes fixed telescopes competitive with tiltable ones. Optical telescopes are seldom used at zenith angles greater than 50 degrees (Benn , & Martin 1987) beyond which the airmass increases rapidly, causing unacceptable degradation of performance from increased absorption and worsening seeing. The BMW is versatile and gives a variety of configurations having different focal lengths and geometries. A mechanically simple configuration (Figure 1) places the tertiary on the optical axis of the primary so that the tertiary-detector assembly looks like a small altazimuth telescope that views the secondary mirror. To view a different region of sky, the secondary would move on a polar coordinates mount while the tertiary would move in an altazimuth mount. In principle the telescope could track by moving and warping the mirrors to follow an object in the sky, but the shapes of the mirrors would have to change in real time. A simpler system can use rigid mirrors with a survey telescope tracking electronically in the TDI mode with a CCD detector (Hickson et al. 1994). The optical and mechanical setups are then considerably simpler since the corrector is set for a particular zenith distance and does not have to be adjusted to work at different zenith distances. A corrector designed for a given zenith distance $\theta$ (e.g. 20 degrees) can be used to observe objects passing anywhere within a field of view of $2\theta$ (e.g. 40 degrees) by moving it at different azimuths but at a fixed zenith distance, as explained by a figure in Borra, Moretto , & Wang (1995). More recent design work (Moretto 1995) gives smaller secondary and tertiary mirrors and more compact telescopes than reported in the exploratory work of Borra, Moretto , & Wang (1995). For example, using secondary and tertiary mirrors warped with Zernike polynomials, at 7.5 degrees from zenith, Moretto (1995) find PSFs comparable to those in Borra, Moretto , & Wang (1995) for a f/4 4-m primary having a 1.1-m diameter secondary and a 0.8-m tertiary. We only have begun exploring designs for correctors working far off-axis. This is a nearly unexplored area of optical design so that there certainly are better designs than those found so far.

Once a sufficiently large accessible field is achieved, a fixed primary yields a more efficient system than a classical tiltable telescope, for a classical telescope can only observe a field at a time, while a fixed primary with several correctors could access many widely separated fields simultaneously. This primary-sharing setup, allowing several research programs to be carried out simultaneously, is particularly attractive for the largest and most expensive telescopes. Two large telescopes, one built in Chile and the second in either Hawaii or the Mexican or United States deserts, could access most of the sky. A BMW corrector is also attractive for a lunar-based telescope since weight is critical for such an instrument. We should expect that a fixed telescope equipped with a BMW corrector should be lighter than a conventional tiltable telescope, especially if the primary is a liquid mirror (Borra 1992).

On the other hand, those correctors can only increase the cost of fixed telescopes and they can only be practical if the instruments are sufficiently cheaper than steerable telescopes. No detailed costing analysis has been done; however it seems reasonable to assume that a setup that moves a mass of a few hundred Kg is cheaper than a setup that moves several hundred tons.

The mirror surfaces needed for a BMW corrector are complex and we must address their feasibility and cost. The cost of a complex optical surface can be kept low if it can be generated by mechanically warping a spherical mirror. Elastic theory (Lemaître , & Wang 1995) shows that we can warp spherical surfaces into complicated shapes by applying stresses at selected points. The resulting surfaces are represented by Zernike polynomials, well-known in optics (Born , & Wolf 1980). We recently have carried out experimental work that shows that it is possible to warp an optical quality steel surface by adding Zernike polynomials up to fourth order included plus 5th order astigmatism (Moretto et al. 1995).

## 4. LIQUID MIRROR TELESCOPES

The surface of a liquid follows an equipotential surface; hence the advantage of liquid optics: We can use the equipotentials to shape precise surfaces. The surface of a spinning liquid takes the shape of a paraboloid that can be used as a reflecting mirror. This idea is old but was never taken seriously because a LMT cannot be tilted and cannot track like a classical telescope and because past unsuccessful attempts gave it a bad reputation. There has been renewed interest in the concept because modern technology now gives us alternate tracking techniques (Borra 1982, Borra 1987, Hickson et al 1994). Optical shop tests (Borra et al. 1992; Borra, Content , & Girard 1993) have shown diffraction limited performance and LMTs have been built and yielded astronomical images and research (Content et al. 1989, Hickson et al. 1994). The status of the liquid mirror project has been recently reviewed (Borra 1995) and I shall only briefly summarize it here. A handful of liquid mirrors have been built and are used for scientific work for astronomy, the atmospheric sciences (lidar: Sica et al. 1995), the space science (space debris: Potter 1994) as well as reference surfaces for optical shop tests (Ninane 1995). Liquid mirrors clearly work, at least to a 3-m diameter, the large mirror built so far (Potter 1994). Figure 2 shows images taken with a CCD and the NASA JSC 3-m LMT (Potter 1994). An image taken with the UBC/Laval 2.7-m LMT is shown in Hickson et al. (1994).

Liquid mirrors are interesting to Astronomy because they promise two main advantages over conventional glass mirrors: They are considerably cheaper and it should be possible to build them to larger diameters. These are precisely the qualities that sections 1 and 2 argue are needed to solve the present crisis in optical astronomy. The outstanding limitation of LMs, comes from the fact that they cannot be tilted, hence cannot track mechanically and cannot be pointed to observe different regions of the sky. Tracking is no longer a problem for imagery and low-resolution spectrophotometry, as demonstrated by Hickson et al. (1994), and should be feasible for higher resolution spectroscopy, as argued by Borra (1987). The previous section argues that advances in corrector design should eventually allow LMTs to access fields comparable to those of classical tiltable telescopes.

Section 5 shows that competitive research is possible even with the more limited fields accessible with classical correctors.

Girard (1995) made a detailed analysis of the costs of the components and materials needed to build a complete 2.7-m diameter liquid mirror as well as the time to build and install it. It is based on costs and times spent making and installing a 2.7-m diameter LMT at the University of Western Ontario (Sica et al., 1995) and is summarized in Table 1. We can see that, even factoring in the cost of manpower at $40/hour, this liquid mirror costs 1 to 2 orders of magnitude less than a conventional high-quality glass mirror and its cell. Note that most of the manpower required is unskilled manpower and that the system is simple enough that it can be built in a typical University shop. Also note that this represents the cost of building a prototype so that a better engineered system would probably be less expensive. Our experience shows that operating costs respect the usual yearly 15% of capital cost operating expenses; although costs may be higher in very remote sites. Note however that survey LMTs can easily be automated, given the simple routine nature of the data taking, thus reducing operating costs. The low operating costs come from the fact that LMTs are very simple instruments, mechanically, electronically and software-wise. Consequently, an LMT can be run with part-time manpower.

The largest mirror built so far has a diameter of 3-m (Potter 1994); there are however no compelling reasons why it should not be possible to build them to diameters as large as 10 meters and perhaps beyond (Borra, Beauchemin, & Lalande 1985, Gibson, & Hickson, 1992, Borra 1995). The only serious threat to large LMs comes from the wind induced by the rotation of the container that could generate wind driven waves. However, thin mercury layers greatly dampen any disturbance in the liquid. We recently have made a 2.5-m mirror that had a layer of mercury only 0.5-mm thick. Damping of disturbances was large and bides well for very large LMTs.

This is a very young technology and there is much room for improvement. Clearly our mechanical setups (Borra et al. 1992) should be better engineered, particularly if one wants to build much larger mirrors. Mercury liquid mirrors work but mercury is heavy and efforts should be made to find lighter reflecting liquids since lighter mirrors need less expensive bearings and containers. We are investigating mirrors that use gallium and its low melting temperature eutectic alloys, that have less than half the density of mercury. Gallium has a relatively high melting temperature ( 30 degrees C) but this is not an obstacle for our experiments show that it is easy to supercool and very stable in the supercooled state. We have supercooled Ga samples to -27 degrees C. We have made a 1-m diameter gallium-indium mirror (Tremblay 1995) with a reasonable surface quality but some minor, albeit stubborn, technical problems are still pending.

It would be highly desirable to find a low density liquid having high reflectivity and high viscosity. An intriguing possibility is to use a low density high-viscosity, but low-reflectivity liquid, and chemically deposit a reflective metal coating on it. Chemical deposition of thin metal coatings on solids works (Stremsdoerfer, Martin, & Clechet, 1994) but still has to be demonstrated on liquids. It may be necessary to proceed in two steps: first depositing a thin solid pellicle than metal coating it. Unpublished interferometry on small samples shows that it is possible to deposit thin high quality pellicles on viscous liquids. It may also be possible to increase the reflectivity with a liquid dielectric stack.

Non-rotating liquids have (nearly) flat surfaces and we know that GRINs and zone plates having flat surfaces can focus light. There thus are at least two types of non-rotating liquid mirrors: a GRIN LM and a zone plate LM. Non-rotating LMs do not generate the wind that will probably set the ultimate limit of rotating LMs. They should have essentially no size limit and may be used to build optical telescopes having gigantic dimensions. A non-rotating liquid mirror based on GRIN (gradient index) optics has been proposed by Borra (1987b). It consists of a stationary container having a thin layer of mercury covered with a transparent liquid in which one introduces a radial dependence of the index of refraction. It focuses light like a mirror but has the chromatic aberrations of a lens. The index gradient could be achieved with a chemical composition gradient set up by a diffusion-based system using membranes. Verge and Borra (1991) found that it would be possible to obtain a satisfactory mirror with existing chemicals; however, they also found that a binary mixture of liquids is unstable against turbulent convection, rendering the mirror useless. They propose to use a ternary liquid mixture to eliminate the density gradient that drives the convection. No working model has been attempted.

Vibrations induce concentric ripples on a stationary pool of mercury. This suggests that a zone plate (Born , & Wolf 1980) may be built by inducing ripples on a flat mercury surface having the appropriate profiles for a zone plate to focus light. It would however have the strong chromatic aberrations typical of waveplates which could however be minimized with holographic correctors. As far as I know, such a system has never been proposed before, never been analyzed theoretically and no working model built. In principle one could remove off-axis aberrations by introducing the proper fringe shapes and spacing.

Among interesting technological developments, one should mention that Shuter and Whitehead (1994) have proposed to use magnetic fields to shape the parabolic primary of a ferrofluid mercury telescope into a sphere. They discuss a corrector that gives correction over a 10X10 degree field, allowing thus observations over nearly 4000 square degrees. Ragazzoni and Marchetti (1994) have shown experimentally that magnetic fields do shape flat mercury mirrors and proposed to use them to shape rotating mirrors. They also have proposed to use magnetic fields to shape large space LMTs (Ragazzoni, Marchetti, & Claudi 1994).

The technology is not mature yet and one must carefully differentiate what it promises from what it can deliver now. In the future, it promises inexpensive large mirrors that can, with innovative correctors such as those discussed in section 4, access large regions of sky to carry out the same type of research done with classical telescopes. LMTs also promise large lunar (Borra 1991) or perhaps even orbiting telescopes (Borra 1992, Ragazzoni, Marchetti, & Claudi 1994). Presently, with classical correctors (1-degree correction) they can be used to carry out surveys. The next section examines some research that could be done with a LMT presently in construction that will have a diameter of 5 meters, only a little larger than the largest LMT built so far (3 meters) and feasible with present technology.

## 5. EXAMPLES OF RESEARCH THAT CAN BE DONE WITH THE PRESENT TECHNOLOGY: COSMOLOGY WITH LMTs

Liquid mirror telescopes promise major advances for deep surveys of the sky and, in particular, cosmological studies. The crisis in optical astronomy discussed in section 2 is particularly acute in cosmology since cosmological objects are faint and, furthermore, cosmological studies tend to be statistical in nature, hence need a large number of objects; and therefore the need for considerable observing time on large telescopes. It is difficult to dedicate a large number of observing nights to a specific project with conventional large expensive telescopes. On the other hand, inexpensive LMTs can be dedicated to a specific project. The outstanding limitation of LMTs, that they can only observe near the zenith, is not a serious handicap for cosmological surveys.

As a practical example, let us consider a LMT carrying out an imaging survey through interference filters with a CCD tracking in the TDI mode. Such a survey is in progress with the 2.7-m diameter UBC-Laval LMT (Hickson et al. 1994) and a second one is planned with a 5-m LMT presently under construction.

### 5.1 The data

The data consists of images from which one can obtain morphologies, accurate positions, magnitudes and spectrophotometry having the wavelength resolution of the filters. Liquid mirrors are diffraction limited (Borra, Content , & Girard 1993) so that observations will be limited by seeing, provided it is greater than the effects discussed by Gibson , & Hickson (1992), which is the case in 1 arcsecond seeing at latitudes smaller than 30 degrees. Tracking is done with the TDI technique which is a driftscan at the sidereal rate so that the data enjoy the accurate flatfielding one obtains driftscanning, and which is needed for accurate photometry, particularly for faint objects near the sky background. The spectrophotometry has somewhat lower resolution that one is accustomed to with astronomical spectrographs but it is comparable to the resolution of the Oke Palomar multichannel photometer (Oke 1969). This was one of the most productive astronomical instruments ever, producing some 300 articles (Oke, private communication), and one can imagine similar data for every object in 400 square degrees of sky.

That the limited field of view of a LMT equipped with a conventional 1-degree corrector is not a serious handicap for cosmology is illustrated by Fig. 3 that shows the number counts /square degree of  quasars (from Hartwick , & Schade 1990) and galaxies (from Lilly, Cowie , & Gardner 1991 and Metcalfe,  Shanks,  Fong, & Jones  1991), brighter than a given B magnitude, at the galactic poles. It predicts that a survey to B= 24 would observe, in a square degree of sky at the galactic poles, 20,000 galaxies and 500 quasars (extrapolating linearly from the last data in Fig. 3). Approaching the galactic plane, the galaxy and quasar counts decrease somewhat because of galactic extinction; but at 60º degrees from the poles we would still have (assuming a cosecant extinction law) 15,000 galaxies/square degree and 250 quasars/square degree. Let us only consider the strip of "extragalactic" sky having galactic latitude > 30 degrees and conservatively assume that the optical corrector of the telescope yields good images over a 1 degree field, well within the performance of existing corrector designs. We would observe 100 to 200 square degrees of " extragalactic", depending on the latitude of the observatory , and, at a latitude of 30 degrees, roughly 3 $10^6$ galaxies and  50,000 quasars with B< 24. For comparison, the center for astrophysics redshift survey  has so far observed 15,000 galaxies to redshifts < 0.05 in over 15 years of operation. Also for comparison, the total number of quasars in the latest quasar catalog (Hewitt, & Burbidge 1993) contains 7,000 objects gathered in 30 years, but is essentially useless for statistical  studies since the objects were identified from a variety of search techniques having poorly quantifiable selection effects.

## 5.2 Redshifts from low resolution spectrophotometry

Redshifts give crucial cosmological information and one may wonder how accurately they can be determined from low resolution spectrophotometry. They can be obtained from cross-correlation techniques that also classify the galaxy by its spectral type. Hickson, Gibson and Callaghan (1994) have carried out  simulations to determine redshifts with the cross-correlation technique applied to noise-degraded SEDs of galaxies. They assume the interference filters of the UBC-Laval survey and obtain, for a signal to noise ratio (S/N)  of 10, an rms redshift error less than 6,000 Km/sec for the mix of all galaxy types, and rms morphological errors less than 0.14. The precision is improved to 3,000 Km/sec  for early Hubble types. Cabanac (1992) reached similar conclusions also on the basis of crosscorrelation techniques.

Cabanac (1992) has also carried out Montecarlo simulations of noise-degraded spectra that were then analyzed with the break-finding algorithm described by Borra and Brousseau (1988). Redshifts were determined from the position of the redshifted 4,000 Å break that is conspicuous in elliptical galaxies but  decreases in later Hubble types. Cabanac finds that at a S/N = 10, the software correctly assigns redshifts for 95% of E and S0 galaxies within 3,000 Km/sec decreasing to 75% for Sab and to 40% for Sbc. The success rates increase with S/N. Cabanac finds that a major source of uncertainty arises because the software has a hard time  distinguishing,  in the presence of noise, the redshifted 4,000Å break from other spectral features. If a reliable algorithm can be found to correctly identify the 4,000 Å break among the others, the precision will increase considerably. For example, at a S/N = 10 the breakfinder yields a standard deviation of 1,000 Km/sec for galaxies as late as Sb, if it correctly identifies the redshifted 4,000 Å break.

In conclusion, Montecarlo simulations indicate that morphological types can be obtained with a reasonable precision. For a S/N at least 10, redshifts can be obtained with, at worse, a rms error of 6,000 Km/sec that can improve to 1,000 km/sec for early type galaxies, a sufficient accuracy for many cosmological studies, in particular studies of the large scale structure of the Universe. One may be skeptical of this kind of redshift accuracy quoted for low resolution and signal to noise ratio spectra; however Beauchemin and Borra (1994) successfully detected redshift peaks corresponding to large scale structure found independently by others. This was accomplished with a somewhat higher resolution (75Å) but with noisy photographic slitless spectra.

## 5.3 Performance of a 5-meter diameter LMT

Recently, a Canadian-French collaboration has obtained funding for a 5-m diameter LMT that will be equipped with a 3072 x 4608 CCD mosaic having 15-micron pixels. It will carry out a spectrophotometric survey, with the 40 interference filters described by Hickson et al. (1994). Let us consider the performance expected of this 5-m LMT tracking with a CCD in the TDI mode. The assumptions and equations used in the computations are described by Borra (1995). The signal to noise ratio of a flux measurement is obtained from photon statistics and includes contributions from the sky, CCD read out noise and dark counts. I use an aperture of 3.0 arcseconds, as in Lilly, Cowie and Gardner (1991) and a sky background of 22.5 magnitudes/square arcsecond at the zenith. I did not include any aperture correction for it is not easy to quantify, since it depends on the type of galaxy, redshift, cosmological model, wavelength, etc... On the other hand, Lilly, Cowie and Gardner (1991) find that, with good seeing, a 3-arcsecond diameter aperture encloses 95% of light for stellar objects and about 80 % even for their largest galaxies. They also find that the deconvolved galaxy images have 50% light within a 1 arcsecond radius, suggesting that the aperture correction is small even for galaxies having B=21.

Figure 4 shows the signal to noise ratio expected for 140 second integration time as a function of blue magnitude for a f/1.5 5-m diameter LMT observing through a 200 Å filter centered at 4400 Å. With a 3072 x 4608 15 micron pixels CCD mosaic, 140 seconds correspond to a single TDI nightly pass for a site at a latitude of 32 degrees. Figure 4 predicts a signal/noise = 20 at B=21.6, =10 at B= 22.4 and = 5 at B = 23.2 for a single nightly pass. The telescope will cover the wavelength region from 4,000 Å to 10,000 Å with 40 interference filters having logarithmically increasing widths and adequate overlap (Hickson et al 1994). As explained in Borra (1995), assuming a complete spectral coverage from 4000 Å to 10,000 Å, taking account losses to weather (in a good site) and technical problems, one would get 3 passes/filter/year for a total integration time of 420 seconds/filter. In 4 years we would get 12 passes for a total integration time of 1,680 seconds/filter. Table 2 gives the performance of the 5-m LMT, it shows that 4 years of observing would get us to almost B=24 with a S/N = 10, sufficient to get 6,000 Km/sec to 1,000 Km/sec redshift errors, and over B=24 with S/N =5, sufficient for rougher energy distributions and redshifts. The increase in sky brightness with wavelength is roughly compensated by the flux increase with wavelength for most faint galaxies, at least for $\lambda <$ 7,000 Å.

The same table gives the performance expected for observations with a 1,000 Å filter: It reaches B= 27 with S/N = 5, comparable to the faintest observations done so far (e.g. Lilly, Cowie and Gardner 1991) that were painstakingly gathered over several years on 4-m class telescopes with a sky coverage of a few square arcminutes and small statistics (a few hundred objects) that only allow a glimpse at the universe at those magnitudes and pale with respect to the nearly 200 square degrees coverage and millions of objects that a 4 year LMT survey would observe. One can get a vivid impression of what the data will look like by examining the photograph of a 200X200 arcseconds CCD frame that reached 28th magnitude (Fischer, Tyson, Bernstein, & Guhathakurta. 1994) . The LMT survey would get similar data over an area of nearly 200 square degrees, giving a huge database that could be used for a multitude of scientific investigations. Variability information can be extracted from nightly observations. The same performance could of course be obtained with a glass mirror; but only LMTs make it practical since their low cost allows one to afford a telescope dedicated to a specific project.

A significant increase in performance (Table 2) will come from a new generations of three-dimensional detectors that measure the energy of the incoming photons as well as their positions ( Keller, Graff, Rosselet, Gschwind, & Wild, 1994; Perryman, Foden, & Peacock 1993); although the technology is still in its infancy. The prospect of carrying out spectrophotometry to B ~ 27 is an exciting one. Note that at one of these systems is capable of extremely high spectral resolution ( Keller, Graff, Rosselet, Gschwind, & Wild, 1994).

To see how these observations would sample the Universe, I have computed the redshift distribution expected from a survey having lower limiting magnitude $m_0$ and upper limiting magnitude $m_1$ from the usual cosmological integral

$$\frac{dN}{dz} = d\Omega \int_{m0}^{m1} \Phi(M) \frac{dV}{dz} dm \quad , \qquad (3)$$

where $\Phi(M)$ is the differential luminosity function (per unit magnitude) of galaxies, M($H_0$, $q_0$, z, m) the absolute magnitude, m the apparent magnitude, $d\Omega$ the surface area element and dV($H_0$, $q_0$, z) the cosmological volume element/$d\Omega$ , $H_0$ the present epoch Hubble constant, $q_O$ the deceleration parameter and z the redshift. I use $H_0$ = 100 km/sec/Mpc and $q_0$=1/2. The models have been computed with the mix of galaxy types and the K-corrections described in Shanks, Stevenson, Fong, & MacGillivray (1984) with the difference that the parameters of the Schechter luminosity function are derived from the CfA survey (De Lapparent, Geller, & Huchra, 1989) . This is an oversimplification since the luminosity function varies with morphology and the Shechter function is an average over all Hubble types. Furthermore, the luminosity function evolves and I neglected evolution, a reasonable but not perfect assumption for the redshift depths involved. The uncertainties brought by these assumptions are tolerable for our purpose since we are only interested in an estimate of the redshift space sampled, rather than detailed modeling. In any event, more sophisticated simulations would rest on assumptions without solid empirical backing. Figure 5 shows the redshift distributions expected for surveys reaching 22nd, 24th and 28th blue magnitudes.

The huge database given by nearly 200 square degrees of images could be used for a multitude of projects. Basically one could use the data for the same kind of research that can be done with a Schmidt telescope but with accurate magnitudes and variability or spectrophotometric information.

5.4 Selected projects

There obviously is a large variety of cosmology that can be done with the kind of data just discussed. Table 3 gives a short list, certainly not inclusive, of topics that could be addressed with an LMT. The "Big Questions" to address change with time, as will the instrumentation used. I briefly examine below some Science of current interest that one could do with the spectrophotometric information gathered with a 5-m LMT.

### 5.4.1 q$_0$ from galaxy counts

As an example of the impact of LMTs in cosmology, let us consider the determination of q$_0$. This is a notoriously difficult measurement since curvature is small at low redshifts and one therefore traditionally has tried to obtain it, with a variety of methods, from observations at high redshifts. A first difficulty arises because, having to observe far, one needs intrinsically bright (or large) objects that tend to be rare. A second, and worse, difficulty is caused by evolution effects, important given the great lookback time at high z, that have bedeviled efforts to get q0: geometry and evolution enter all tests and cannot be disentangled.

Volume tests give the most sensitive measurements for q$_0$. Consider for example, the number of objects /unit surface/unit redshift, it is given by

$$\frac{dN}{dz} = d\Omega N_0 c^3 [q_0 z + (q_0 - 1)(\sqrt{1 + 2q_0 z} - 1)]^2 / [H_0^3 (1 + z)^3 q_0^4 \sqrt{1 + 2q_0 z}], \qquad (4)$$

where the symbols have the usual meaning and $N_0$ is the space density at z=0.
As with all geometrical tests, the difference among the various geometries only becomes large for z >1, where only intrinsically bright rare objects are detectable (e.g. QSOs) and where evolution effects are large; but is small for z<0.3 where evolution is much smaller and where intrinsically fainter and more numerous objects (e.g. ordinary galaxies) are detectable. The difficulty is illustrated by the 3 curves in figure 6 that shows the counts predicted at z = 0.1, 0.3 and 0.5 as function of q$_0$, normalized to the counts predicted for q0=0.5. Let us determine whether we live in an open or closed Universe and take the criterion that we can differentiate between q$_0$ = 0.4 and q$_0$ = 0.5 at the 5 σ level. At z = 0.3, the ratio between the counts for q$_0$ = 0.4 and for q$_0$ = 0.5 is 1.05. If we ask for a 5 standard deviation discrimination and assuming that H$_0$ and N$_0$ are known, Poisson statistics demand a minimum of 10,000 objects. Of course Poisson statistics cannot be blindly applied because galaxies are known to be clustered on a scale of a few Mpc; there will be an excess of variance in the cell counts compared to a random distribution. However, the correction factor that takes into account the departure from Poisson statistics scales as 1/volume so that the correction is negligible for the large volume of sky that we sample, provided galaxies remain clustered at z=0.3 as they are in the local universe. This can be checked with the data

Figure 5 shows that there are 1.2 10$^3$ objects/square degree with 0.25 <z<0.35 and B< 24. A telescope located at a latitude of 30 degrees observes a one-degree wide extragalactic strip of sky (bII<30º) containing 175 degrees . Allowing for galactic extinction, the telescope could observe 200,000 galaxies with 0.25 <z<0.35 in the one-degree-wide strip accessible with a conventional corrector. It will not be possible to obtain redshifts for all objects and there will also be some loss due to some large nearby galaxies and the halos of bright stars. However, there will be plenty of ordinary elliptical and spirals to get well over 10,000 redshifts. Note that over 1/3 of the galaxies should be ellipticals and early type spirals for which σ ~ 1000 Km/sec can be obtained from the 4,000 Å break (Cabanac 1992).

In practice, one would have to obtain counts at various redshifts and use a least-squares fitting procedure and he would have to consider systematic effects that may cause spurious z-dependent gains or losses of objects (e.g. magnitude cutoff, redshift or photometric errors). The large quantity of data should leave us well equipped to understand this. For example, there is some evidence that the faint end of the luminosity function evolves at surprisingly low redshifts. The data would allow to measure the evolution and either correct for it or simply truncate the luminosity function at the appropriate magnitude.

Close attention shall have to be paid to effects peculiar to the data, such as the effect of the large redshift errors which depend on the Hubble type. This is not the place to carry out a detailed discussion; the relevant point is that the data allows us to contemplate such a project at all. The theme of subsection 5.4 is that having a large telescope dedicated to a project allows us to consider a project that would be otherwise unthinkable.

### 5.4.2 Large Scale structure

Figure 5 shows that the universe is sampled to significantly higher redshifts than any other large scale redshift existing (e.g.CfA) or planned (e.g. the Sloan Digital Survey), albeit with a considerably lower radial velocity precision. One could therefore study the large scale structure of the universe from a few times the radial velocity precision to the redshift depth of the survey (a few Gpc). This complements the information that will be obtained from the other more precise but shallower surveys. Figure 7 (Adapted from Vogeley 1995) shows the uncertainty of the SDSS power spectrum. The $1\,\sigma$ uncertainty expected for a volume limited (to M*) sample of the SDSS northern redshift survey, assuming Gaussian fluctuations and a $\Omega h=0.3$ CDM model, is compared to power spectra for CDM with different $\Omega h$. Power bars on smaller scales are of similar or smaller size than the symbols. The rectangular box shows the range of the HPBW of the z distribution of the 5-m survey (see Fig. 5). The HPBW of the 5-m survey extends well beyond the HPBW sampled by the SDSS. Because the total numbers of galaxies are similar in the 2 surveys, we can expect a similar distribution of error bars. The error bars in the box will approximately have the sizes of the error bars of the SDSS for $\lambda<200$ MPc. We can see that the 5-m survey samples, with small error bars, the very long wavelengths at which the differentations among the theoretical models is the greatest, and overlaps with the scales probed by COBE. Hopefully, the small error bars given by the large statistics may be able to detect the features predicted by some models; if needed the statistics can be increased by observing different strips of sky. The telescope could be moved or other ones build, an acceptable alternative given the low cost of the system.

Because we get energy distributions, morphologies and accurate photometry we can repeat the analysis as function of Hubble type, spectral type, etc.... We also can study the redshift dependences of the energy distributions and mixes of Hubble types.

### 5.4.3 Gravitational lenses

Gravitational lensing has become a very active field of research since the discovery of the doubly imaged quasar Q0957+561 AB (Walsh, Carswell , & Weynman 1979). There is now a very large literature on the subject and a growing number of applications of gravitational lensing (Refsdal , & Surdej 1994). Reading the review by Refsdal , & Surdej (1994) one sees that Liquid Mirror Telescopes can be used for a variety of research projects in gravitational lensing (e.g. finding new lenses, monitoring brightness variations, etc...). Arguably, mapping the mass distribution in clusters of galaxies, and thus the dark mass distribution, from the distortion introduced by the cluster on the images of background galaxies is the most interesting cosmological application of lensing. The number of clusters mapped so far is small, given the challenge of identifying high redshift clusters and, especially, imaging the fields to very faint magnitudes.

We can estimate the number of clusters observed in the area of sky A of the 5-m survey from

$$N = n(>M)A \int_{0.1}^{z} \frac{dV}{dz} dz , \qquad (5)$$

where *dV/dz* has the same meaning as in Eq. 4 and *n(>M)* is the space density of clusters of galaxies having mass greater than *M*. We use the mass function of clusters of galaxies *n(>M)* determined by Bahcall and Cen (1993). Table 4 gives the number of clusters observable in the 175 square degree strip of extragalactic sky as function of z and richness class. We can see a significant number of clusters for which we can determine the distribution of mass, although we will loose part of the cluster for those at the edges of the field. Figure 5 shows that there will be a sufficiently large number of images of galaxies with z > z(cluster) to be able to do the mapping. At z=0.5, the blue magnitude of the third ranked galaxy B3 ~ 23 so that about 30 galaxies should be identifiable to B3+2.0 for R=0 clusters with z<0.5, and more for richer clusters. Bahcall and Cen (1993) give the relations between *n(>M)*, richness class, and number of galaxies in a cluster with B<B3+2.0. The redshift information will help both to identify the clusters and remove the noise introduced by the images of galaxies with z(galaxy) < z (cluster). The redshift information could also be used to remove the noise of foreground galaxies and thus help in identifying distant clusters beyond the limit at which redshift information can be reliably determined. Given the depth of the observations, the data is also well suited to measure image distortions introduced by large scale structures.

All three projects utilize the massive quantity of data obtained from a large telescope working full time on a project. They were selected from outstanding current problems in cosmological research but the list is certainly not exhaustive.

## 6. CONCLUSION

There is a crisis in optical astronomy caused by the high costs of astronomical telescopes. As a consequence, most astronomers are cut off from competitive research and the rest take an inordinate amount of time to carry out substantial research projects. There is also a heavy price exacted by the inevitable conservatism in the selection of successful research proposals for telescope time. This crisis can be solved with a paradigm shift by developing innovative technologies that will produce inexpensive large telescopes. Using fixed telescopes gives a large saving, albeit at the expense of reduced sky coverage. However, recent work on innovative correctors indicates that fixed telescopes should be able to access surprisingly large regions of sky (as large as 45X360 degrees wide strips). Using Liquid Mirror Telescopes (LMTs) will give another large cost reduction (1 to 2 orders of magnitude). The technology is still in its infancy but optical shop tests of a 2.5-m diameter mirror show diffraction limited performance (Borra, Content , & Girard 1993). Several LMTs having diameters as large as 3-m have been built (Potter, 1994) and gave seeing limited images (e.g. Hickson et al. 1994).

We must carefully distinguish between what can be done now and what might be done later. Fields of views are small with present correctors, so that it is only possible to access a few hundred square degrees of sky with a single LMT; although LMTs located at different terrestrial latitudes will observe different regions of the sky. Tracking has only been demonstrated with the TDI technique so that LMTs can presently only be used as imagers to carry out surveys with filters. Using interference filters one can obtain low resolution spectrophotometry. The type of research that can be done is therefore the type of research that can be done with an imaging survey.

The advantage of the zenithal LMT comes from low cost so that a 4-m class telescope can be dedicated to a narrowly specific project and a huge quantity of data can be generated. I gave three specific examples of LMTs applied to cosmological research; however it should be evident that one can extend similar analyses to any field of Astronomy that uses surveys. The low cost of LMTs is an asset for any such project that needs a dedicated medium to large telescope. Arguably, the more interesting contribution of LMTs may come from serendipity. The history of Astronomy tells us that whenever radically new instruments were devised (e.g. radio-telescopes) some totally unexpected major discoveries were made (e.g. quasars, pulsars). Optical astronomy is the oldest of sciences and the telescope was invented centuries ago; however this will be the first time that massive quantities of data will be generated from large telescopes, and computing power is available to analyze them. In a sense, we can say that we are about to step across a new astronomical frontier: the statistical frontier.

Table 1.
COSTS OF COMPONENTS AND LABOR NEEDED TO CONSTRUCT A 2.7-M
LIQUID MIRROR.
(adapted from Girard, 1995).

=======================================================

| Item | Cost of components (1994 $US) | Labor costs ($40/hour) (1994 $US) |
|------|------|------|
| Complete mirror[1] | 15,400 | 11,600 |
| Safety equipment[2] | 3,800 | 5,600 |
| Installation[3] | 500 | 320 |
| Total | 22,700 | 20,400 |

[1] Complete system, including base, mercury, motor, compressor, etc...
[2] Includes mercury sniffer, safety brakes and anti-spill wheels
[3] In-situ installation, including balancing, debugging, checking image quality

Table 2
PERFORMANCE OF THE 5-M LMT (BLUE MAGNITUDE AS FUNCTION OF
VARIOUS FACTORS).
==========================================================

| Duration of the survey | Narrow-band filters (200Å centered at 4400Å) | |
| | S/N=10 | S/N= 5 |
| --- | --- | --- |
| 1 night, 1 pass (140 seconds) | 22.4 | 23.2 |
| 1 year, 3 passes (420 seconds) | 23.0 | 23.8 |
| 4 years, 12 passes (1,680 seconds) | 23.7 | 24.5 |

Wide band FILTER(1000 Å)

| 1 night | 24.1 |
| --- | --- |
| 1 year, 60 nights | 26.3 |
| 4 years, 240 nights | 27.10 |

Hypothetical three-dimensional imager (200 Å bandpass)

| | S/N = 10 | S/N = 5 |
| --- | --- | --- |
| 1 year, 60 nights, | 24.6 | 25.4 |
| 4 years, 240 nights, | 25.4 | 26.2 |

Table 3
SELECTED COSMOLOGICAL TOPICS THAT CAN BE STUDIED WITH AN LMT
===========================================================
Preselection of objects for further studies (e.g. identification of distant clusters of galaxies
or quasars with absorption systems).
Evolution of the luminosity function of galaxies
Evolution of galaxies
Large scale structure of the universe (with redshifts and/or positions)
Supernova rates
Search for gravitational lenses
Gravitational potentials of clusters and large scale structures with arcs and arclets.
Variable objects (e.g. QSOs)
Fluctuations of sky background to get distant clusters.
Find QSOs and AGNs by variety of techniques (astrometry, variability, spectra, colors)
Search and study of primeval galaxies
Serendipity (This is arguably the most promising)
___________________________________________________________

Table 4
NUMBER OF CLUSTERS IN THE 175 SQUARE DEGREE STRIP OF
EXTRAGALACTIC SKY AS FUNCTION OF z AND RICHNESS CLASS.
==========================================================

| R | $0.1<z<0.3$ | $0.1<z<0.4$ | $0.1<z<0.5$ |
|---|---|---|---|
| $\geq 0$ | 91 | 187 | 315 |
| $\geq 1$ | 40 | 82 | 139 |
| $\geq 2$ | 8 | 16 | 28 |
| $\geq 3$ | 1 | 2 | 3 |

FIGURE CAPTIONS

Figure 1:
Schematic design of a 4-m primary equipped with a BMW corrector observing at 15 degrees from the zenith. The scale is given in the figure. To view a different region of sky, the secondary would move on a polar coordinates mount while the tertiary would move in an altazimuth mount .

Figure 2
 Observations of a 20X20 arcminute field taken with a CCD and the NASA Johnson Spaceflight Center 3-meter LMT. The observations were taken with 0.62 arcseconds pixels and FWHM=2.5 arcseconds.

 Figure 3
Number counts of  quasars and galaxies /square degrees, brighter than a given blue magnitude, at the galactic poles.

 Figure 4
Signal to noise ratio expected for 140 second integration time (a single nightly pass) as a function of blue magnitude for a f/2 5-m diameter LMT observing through a 200 Å filter centered at 4400 Å.

Figure 5
Redshift distributions expected for surveys reaching 22nd, 24th  and 28th blue magnitudes.

Figure 6
The 3 curves in figure  show the galaxy counts  predicted at z = 0.1, 0.3 and 0.5 as function of $q_0$, normalized to the counts predicted for $q_0$=0.5.

 Figure 7
 (Adapted from Vogeley 1995) shows the uncertainty of the SDSS power spectrum. The 1 σ uncertainty expected for a volume limited (to M*) sample of the SDSS northern redshift survey, assuming Gaussian fluctuations and a $\Omega h$=0.3 CDM model, is compared to power spectra for CDM with different $\Omega h$. Power bars on smaller scales are of similar or smaller size than the symbols. The rectangular box shows the range of the HPBW of the z distribution   of the 5-m survey. See text for more details.

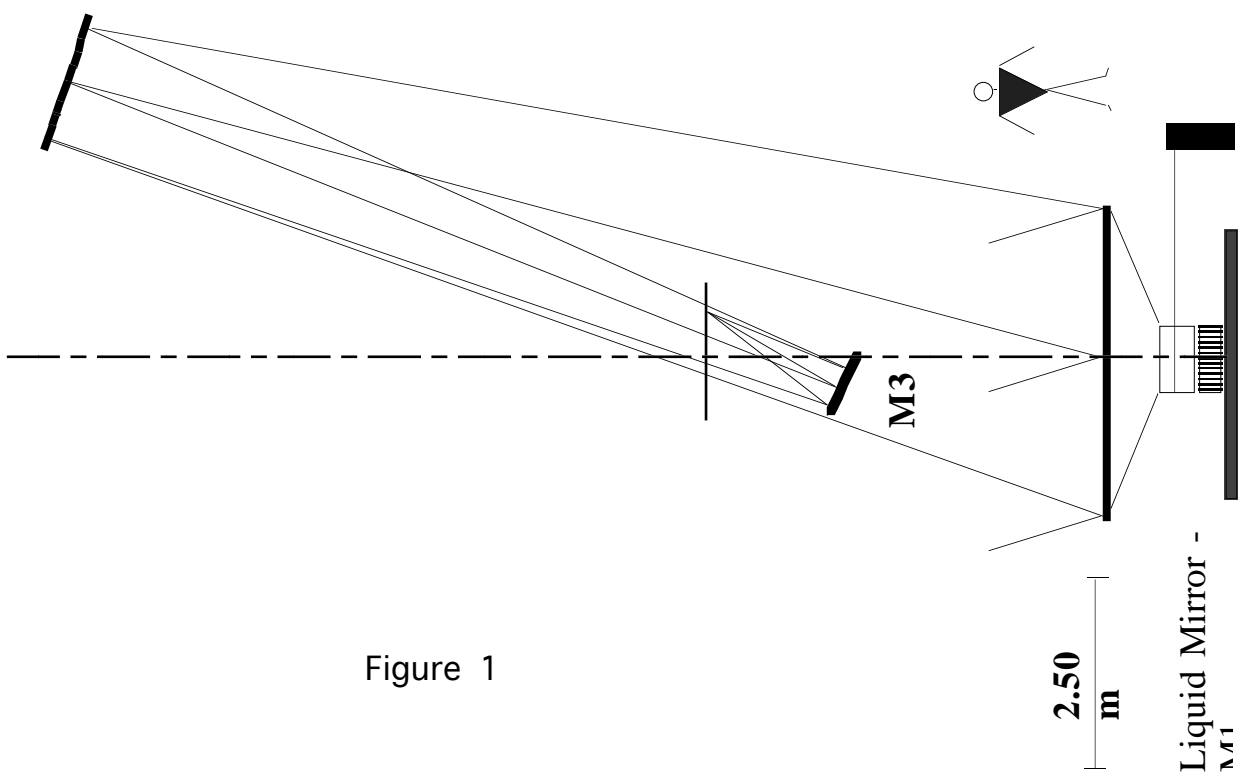

Figure 1

2.50 m

Liquid Mirror - M1

M3

Figure 3

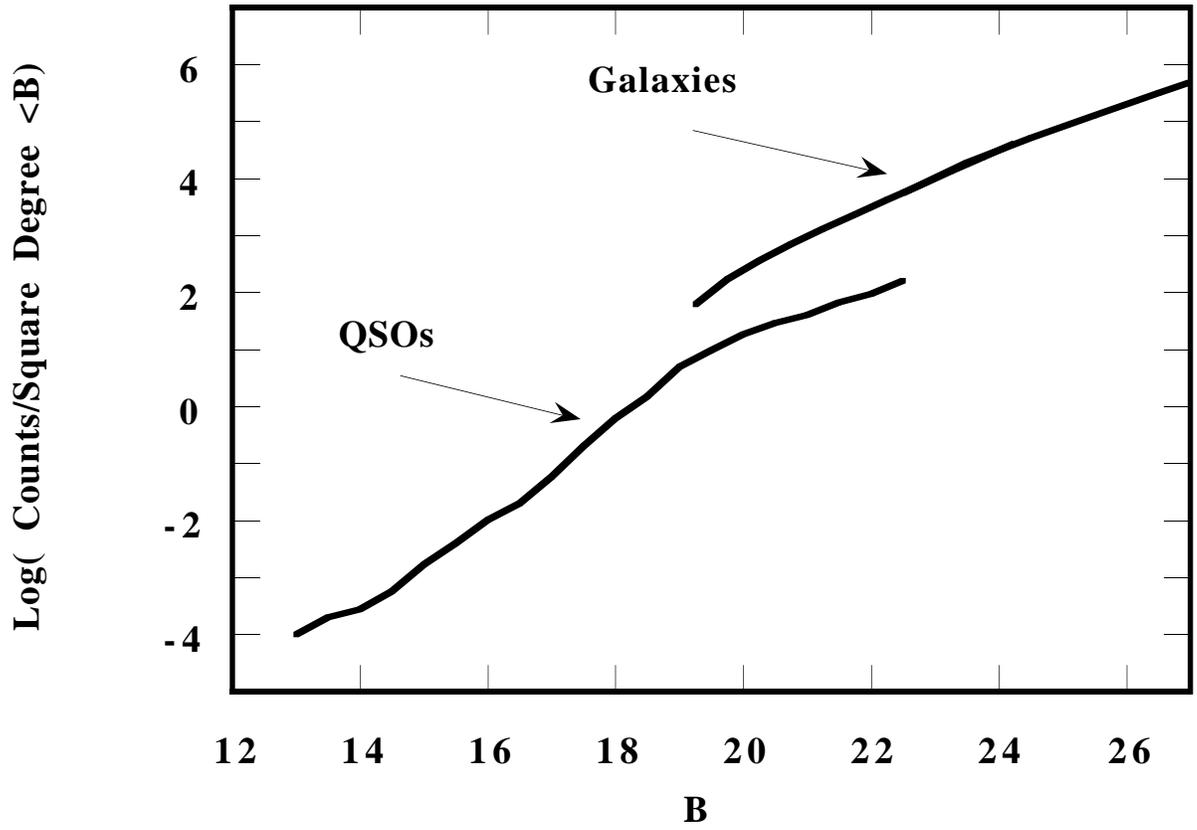

Figure 4

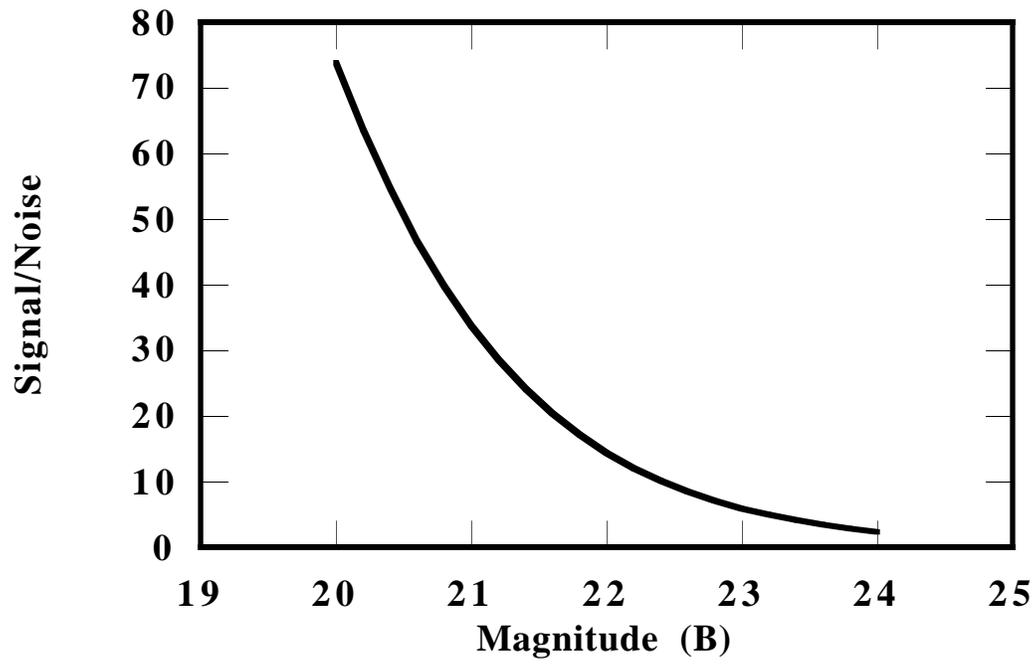

Figure 5

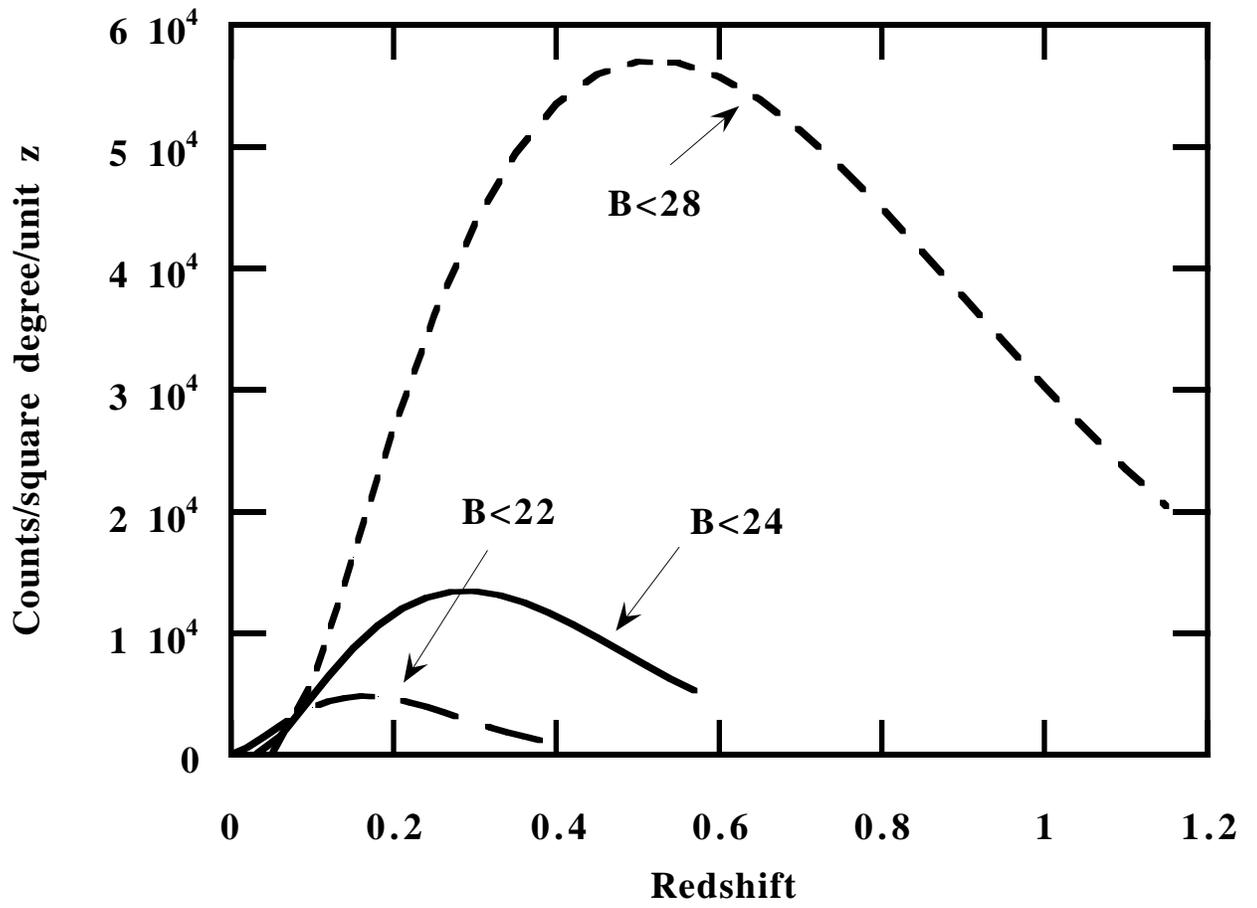

Figure 6

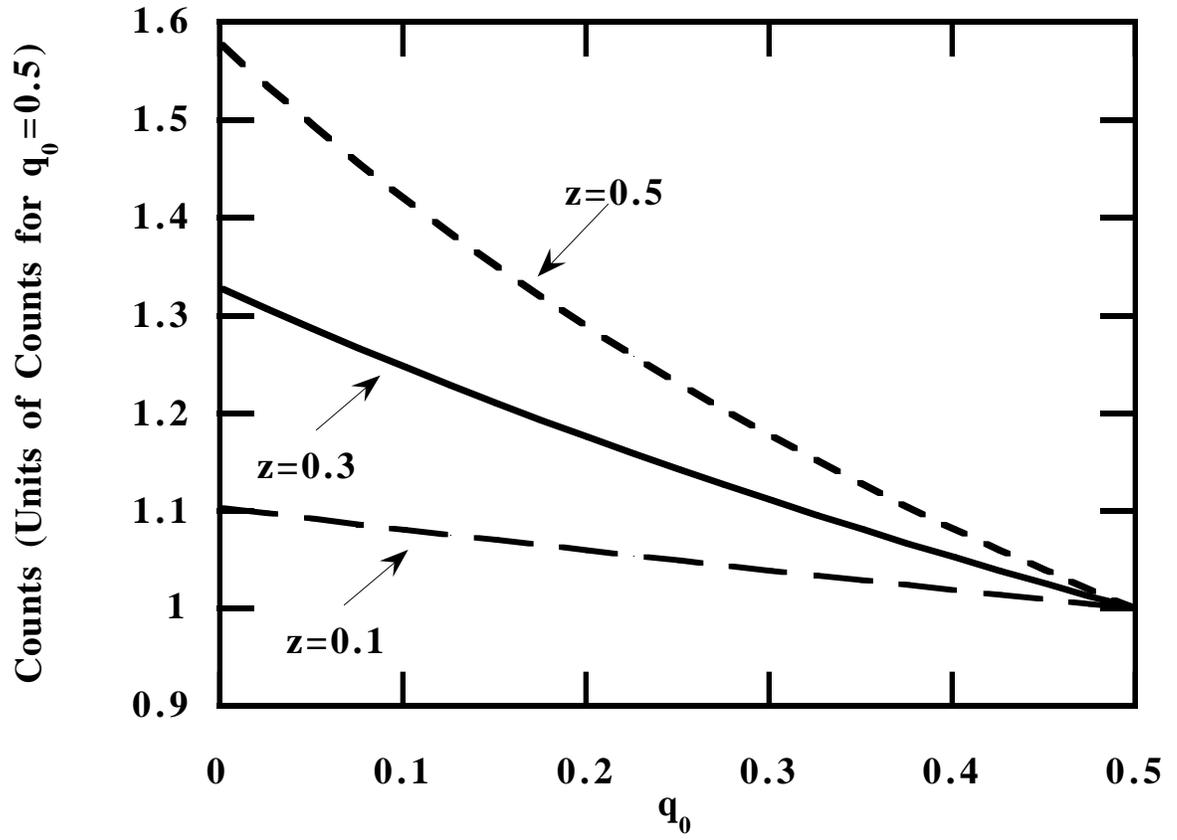

Figure 7

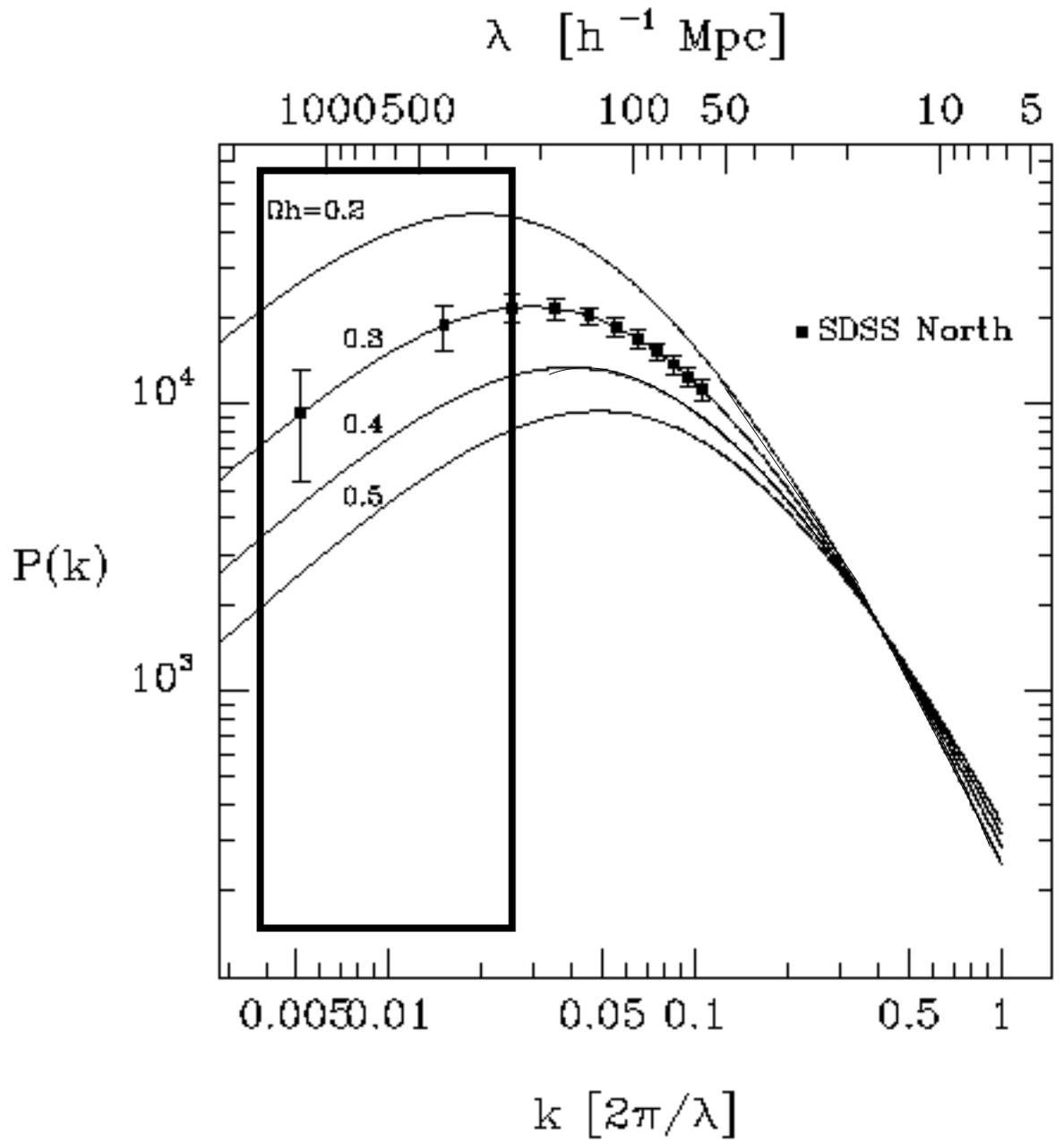